# Communication Language Specifications For Digital Ecosystems

Youssef Bassil

LACSC – Lebanese Association for Computational Sciences
Registered under No. 957, 2011, Beirut, Lebanon
youssef.bassil@lacsc.org

*Abstract:* Service-based IT infrastructures are today's trend and the future for every enterprise willing to support dynamic and agile business to contend with the ever changing e-demands and requirements. A digital ecosystem is an emerging business IT model for developing agile e-enterprises made out of self-adaptable, self-manageable, self-organizing, and sustainable service components. This paper defines the specifications of a communication language for exchanging data between connecting entities in digital ecosystems. It is called ECL short for Ecosystem Communication Language and is based on XML to format its request and response messages. An ECU short for Ecosystem Communication Unit is also presented which interprets, validates, parses ECL messages and routes them to their destination entities. ECL is open and provides transparent, portable, and interoperable communication between the different heterogeneous distributed components to send requests, and receive responses from each other, regardless of their incompatible protocols, standards, and technologies. As future research, digital signature for ECL is to be investigated so as to deliver data integrity as well as message authenticity for the digital ecosystem.

## I. INTRODUCTION

Recently, the business use of the Internet has changed from conception to an ever-present practice. It is no more related with shopping on the web, but transforming business processes into online e-models for performance, scalability, and client responsiveness. This so called e-transformation is a must for every enterprise to prosper in today's digital economy [1]. A digital ecosystem is a distributed IT infrastructure built using interrelated e-service models and aimed at creating an across-enterprise computing environments [2]. Characteristically, a digital ecosystem has such properties as sustainability, standardization, self-organization, self-integration, and self-adaptation [3, 4]. It is inspired by natural ecosystems that evolve and adapt according to their living environment. In fact, digital ecosystems are still in their early development phase in that no clear standards are defined yet [5]. Particularly, there exist no language or protocol specification to formally define the communication and data exchange between the different services of the ecosystem.

This paper proposes a communication language called ECL short for Ecosystem Communication Language. It is a proprietary XML-based language that allows the transparent collaboration and interoperability between different services, possibly built using different architectures, programming languages, and technologies. ECL is powered by the ECU short for Ecosystem Communication Unit which houses the ECL interpreter and delivers a universal data-path to connect services, possibly incompatible, together to interact, send requests, and receive responses from each other.

## II. RELATED WORK

Little work has been done to develop a standard communication protocol for digital ecosystems. A sole attempt is the OASIS reference model [6] which is a generic framework for building and managing service-oriented architectures. It is majorly composed of six units: the orchestration and management unit which is responsible for administering the connected components and web services in the SOA; the data content unit which represents a set of databases that feed web services with data and information;

the service description unit which defines the functions exposed by the connected web services in the SOA; the service discovery unit which contains a look-up registry to locate and consume web services; the messaging unit which can be thought as the communication medium that lets all connected components share data and communicate between each other; and the security and access unit which provides a security layer for securing and encrypting the messages being sent and received between the different components of the SOA. Figure 1 depicts the six units of the OASIS reference model.

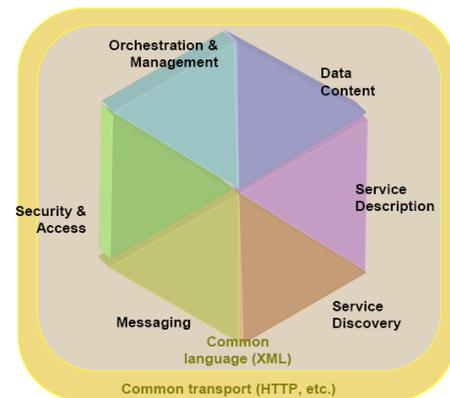

Figure 1.   OASIS model

As for the OASIS message exchange protocol, it defines service participants and their delegates which interact with each other to achieve a particular operation. There are two main modes of interaction: joint actions that cause real world effects and notification of events that report real world effects.

A message exchange is used to execute an action directed towards a specific node with respect to the action model and delegates which are responsible for interpreting the message properly. Furthermore, a message exchange is used to transfer event notifications. An event is an incident that is relevant to some participant. Both action and notification messages have formatting requirements that must comply with the syntax and semantics of the original OASIS specification. The syntax and semantics of the message payload communicated between different parties of the system are also defined by OASIS. It indicates the size, format, and different parameters of every transmitted message. It also defines the exception conditions and error handling for the message payload. When an action needs to be invoked, the correct interpretation must be performed on the receiver's end which itself invokes a set of internal actions that, in turn, drive a particular algorithm or procedure.

## III. ECU - ECOSYSTEM COMMUNICATION UNIT

The proposed Ecosystem Communication Unit (ECU) is a multi-agent and a multi-platform design for connecting services, possibly incompatible, together to interact, send requests, and receive responses from each other. As a multi-agent model, it permits the distribution of services over different machines, networks, and premises, allowing a seamless and transparent communication between them. As a multi-platform model, it permits the support of incompatible services built using different platforms, different standards, different technologies, and different programming languages. Figure 2 depicts the proposed ECU together with its inner-workings.

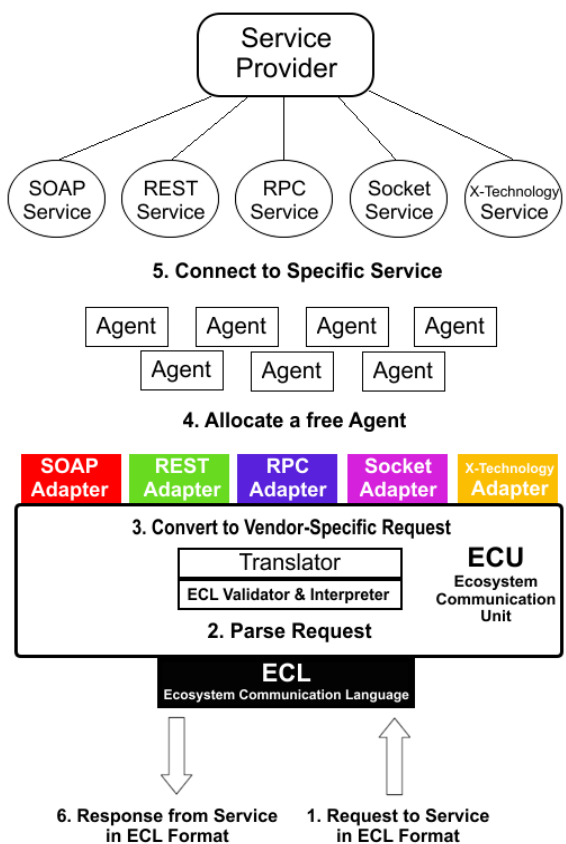

Figure 2.  ECU

In the proposed ECU model, the service provider is the party responsible for providing services to the clients of the existing digital ecosystem. Examples of these services include data storage and retrieval, data feed, ISBN lookup, currency and foreign exchange, and e-commerce. A service is the core component for delivering functionalities to end users and client's applications in the ecosystem. Often, users rely on communication and network setups to connect to remote services and request certain operations. Stereotypically, services are implemented as Web Services or WS for short which are software components that support interoperable machine-to-machine interaction over a network [7]. They additionally provide a WSDL (Web Services Description Language) to describe their offered operations and methods and a UDDI (Universal Description, Discovery and Integration) registry in which they are listed and registered [8]. Practically, web services have several common types, each built using a particular technology and protocol and they are: SOAP (Simple Object Access Protocol) which exchanges data formatted using an XML-based protocol called SOAP; REST (Representational State Transfer) which exchanges data using the different request methods of the HTTP protocol; RPC (Remote Procedure Call) which uses an inter-process communication to call a distributed function located in a remote application; and Socket-based which uses raw TCP or UDP protocols to send and receive data over the TCP\IP protocol.

The agents ensure a distributed ecosystem all the time as they are in charge of handling client's requests to the exiting services. Furthermore, they work as load balancers that distribute workload across multiple servers to achieve optimal resource utilization, maximize throughput, increase performance, minimize response time, and avoid overload. Once a request to a service is received, it is allocated to a free agent that, in turn, allocates it in a round robin fashion to the appropriate back-end machine to process the request. Agents constantly go on and off as requests flow throughout the ecosystem.

Adapters are end-point connectors that bridge a client's request to its destination service. They provide standardization and interoperability as they permit the interaction between two incompatible entities in the ecosystem. The role of the ECU is to identify the type of the request and to route it accordingly to the corresponding adapter which, successively, passes it to the corresponding service.

The Ecosystem Communication Language (ECL) is an XML-based language used to format messages interchanged between the different entities of the ecosystem. The ECL uses XML tags to specify data and their metadata. Besides, an interpreter is used to parse and decode a given ECL message. Another module namely the ECL translator converts a received ECL message into another format compatible with the protocol of the invoked service. In other terms, all ECL request messages sent by clients, whatever their protocols, are first converted to the protocol of their destination services and then processed. As a result, the ECL provides a transparent communication between the different components of the ecosystem allowing them to interoperate despite their incompatible technologies and platforms.

### A.  The Communication Process

Step 1: Client A sends a request in ECL format to a SOAP-based web service B to invoke a certain operation. The ECL request encapsulates metadata describing the request message, including the source client, the destination service, a time stamp, and the function to call.

Step 2: The ECU receives the request message; it first validates the correctness of its XML structure using a DTD document. If it is valid, the request is converted from ECL format into the protocol of service B, in this case SOAP, using the ECL translator. The ECU uses an internal registry to lookup the technical details about service B.

Step 3: The ECU routes the converted request to the adapter that is compatible with the service B, in this case, the SOAP adapter.

Step 4: The adapter then locates a free agent to handle this request. Intrinsically, the free agent tries in sequence to locate the best machine on the network to process the request.

Step 5: Service B gets bound temporary to client A and starts executing the requested function.

Step 6: Once processing is done, a response is sent back from service B to client A. It is first sent to the corresponding adapter, in this case, the SOAP adapter, then to the ECU, then translated to an ECL format, and eventually routed to client A.

## IV.  ECL SPECIFICATIONS

The proposed Ecosystem Communication Language (ECL) is a language specification for exchanging structured data between interconnected services in a digital ecosystem. It relies on XML [9] to format messages which are composed of XML markup tags, denoting the message payload along with its metadata. The metadata denote non-functional information essential for the correct routing and execution of messages within an ecosystem infrastructure. There exist two types of ECL messages: request and response messages, each having a metadata and payload section.

The metadata of an ECL request starts with the *<protocol>* tag indicating a protocol type message. It has several children: The *<sourceIP>* tag which denotes the IP of the entity that initiated the request namely a client, the *<destinationIP>* tag which denotes the IP of the destination entity namely a service, the *<sourceID>* tag which denotes the identification number of the entity that initiated the request, and the *<destinationID>* tag which denotes the identification number of the receiving service. Additionally, the metadata comprise a *<stamp>* tag which denotes the date and time of the request, and a *<version>* tag which denotes the actual version of the ECL protocol.

On the other hand, the payload of an ECL request contains a *<funtionInvoked>* tag which denotes the function to be executed on the destination service, and the *<functionParams>* tag which denotes the list of parameters for the invoked function. The *<functionParams>* tag can have zero or more sets of three children: the *<name>* tag which denotes the name of the sent parameter, the *<value>* tag which denotes the value of the sent parameter, and the *<type>* tag which denotes the data type of the sent parameter. Below is a sample of an ECL request message.

```
<protocol>
    <sourceIP>192.168.1.20</sourceIP>
    <destinationIP>192.168.1.177</destinationIP>
    <sourceID>24</sourceID>
    <destinationID>91</destinationID>
    <functionInvoked>Max</functionInvoked>
    <functionParams>
        <name>x</name>
        <value>10</value>
        <type>int</type>
        <name>y</name>
        <value>15</value>
        <type>int</type>
```

```
    </functionParams>
    <stamp>11/4/2011 09:32:10PM</stamp>
    <version>1.0</version>
</protocol>
```

In contrast, the metadata of an ECL response starts with the *<protocol>* tag indicating a protocol type message. The protocol tag has several children: The *<sourceIP>* tag which denotes the IP of the entity that is returning back the response namely a service, the *<destinationIP>* tag which denotes the IP of the entity that originally initiated the request namely a client, the *<sourceID>* tag which denotes the ID of the service sending the response, and the *<destinationID>* tag which denotes the ID of the receiving client. There also exist the *<stamp>* tag which denotes the date and time of the request, and the *<version>* tag which denotes the actual version of the ECL protocol.

The payload of an ECL response contains a *<returnValue>* tag which denotes the actual value returned by the service and the *<returnType>* tag which denotes the data type of the returned value. Below is a sample of an ECL response message.

```
<protocol>
    <sourceIP>192.168.1.177</sourceIP>
    <destinationIP>192.168.1.20</destinationIP>
    <sourceID>91</sourceID>
    <destinationID>24</destinationID>
    <returnValue>15</returnValue>
    <returnType>int</returnType>
    <stamp>11/4/2011 09:32:13PM</stamp>
    <version>1.0</version>
</protocol>
```

## V.  ECL VALIDATION

In order to validate whether an ECL request or response conforms to the specifications of the ECL language, the ECU employs a DTD validator [10] which verifies the correctness of the grammar and syntax of an ECL message. Below are two DTD definitions: the first one is to validate an ECL request type message, while the second is to validate an ECL response type message.

```
<!ELEMENT protocol (sourceIP, destinationIP, sourceID,
destinationID, functionInvoked, functionParams, stamp, version)>
<!ELEMENT sourceIP (#PCDATA)>
<!ELEMENT destinationIP (#PCDATA)>
<!ELEMENT sourceID (#PCDATA)>
<!ELEMENT destinationID (#PCDATA)>
<!ELEMENT functionInvoked (#PCDATA)>
<!ELEMENT functionParams ( (name, value, type)* )>
<!ELEMENT name (#PCDATA)>
<!ELEMENT value (#PCDATA)>
<!ELEMENT type (int|double|string|int[]|double[]|string[])>
<!ELEMENT stamp (#PCDATA)>
<!ELEMENT version (#PCDATA)>

<!ELEMENT protocol (sourceIP, destinationIP, sourceID,
destinationID, returnValue, returnType, stamp, version)>
<!ELEMENT sourceIP (#PCDATA)>
<!ELEMENT destinationIP (#PCDATA)>
<!ELEMENT sourceID (#PCDATA)>
<!ELEMENT destinationID (#PCDATA)>
<!ELEMENT returnValue (#PCDATA)>
<!ELEMENT returnType (int|double|string)>
```

```
<!ELEMENT stamp (#PCDATA)>
<!ELEMENT version (#PCDATA)>
```

## VI.   ECL ENCRYPTION

The ECU uses the XML Encryption standard also known as XML-Enc [11] to encrypt the payload of an ECL message. Using XML Encryption, structured data can be exchanged in a secure and safe way. In fact, only the payload of an ECL message is encrypted while leaving the metadata unencrypted. This includes the <*functionInvoked*> and the <*functionParams*> tags of an ECL request, and the <*returnValue*> and the <*returnType*> tags of an ECL response. The cryptography algorithm used to encrypt and decrypt ECL messages is the Triple-DES originally published in 1998 [12] which applies the classical Data Encryption Standard (DES) with a 56-bit key three times on 64-bit data blocks. A sample of an ECL request message with an XML encrypted payload is shown below:

```
<protocol>
    <sourceIP>192.168.1.20</sourceIP>
    <destinationIP>192.168.1.177</destinationIP>
    <sourceID>24</sourceID>
    <destinationID>91</destinationID>
    <EncryptedData
        Type='http://www.w3.org/2001/04/xmlenc#Element'
            xmlns='http://www.w3.org/2001/04/xmlenc#'>
            <CipherData>
                <CipherValue>xpQM3w/ah/kpUfCizLu9j…
                <CipherValue>
            </CipherData>
    </EncryptedData>
    <EncryptedData
        Type='http://www.w3.org/2001/04/xmlenc#Element'
            xmlns='http://www.w3.org/2001/04/xmlenc#'>
            <CipherData>
                <CipherValue>xpQM3w/ah/mkH/YDkUBXQP…
                <CipherValue>
            </CipherData>
    </EncryptedData>
    <stamp>5/4/2011 09:32:10PM</stamp>
    <version>1.0</version>
</protocol>
```

## VII.   EXPERIMENTS & RESULTS

In the experiments, an E-learning digital ecosystem model was tested. It comprises three layers: The presentation layer delivering the system's input and output interfaces; the service layer hosting all the system's services; and the data layer housing the system's data storage. The service layer is majorly composed of several web services ready to be consumed by users and client applications. They are but not limited to the "QUIZ" web service which issues exams and workouts; the "TUTORIAL" web service which represents a virtual interactive tutor; the "ENCYCLOPEDIA" web service which offers articles and extracts; the "DICTIONARY" web service which provides a look-up service for finding the meaning of words; and the "SEARCH" web service which helps students find documents, articles, handouts, and other learning materials. Figure 3 shows the diagram of the E-learning model under test.

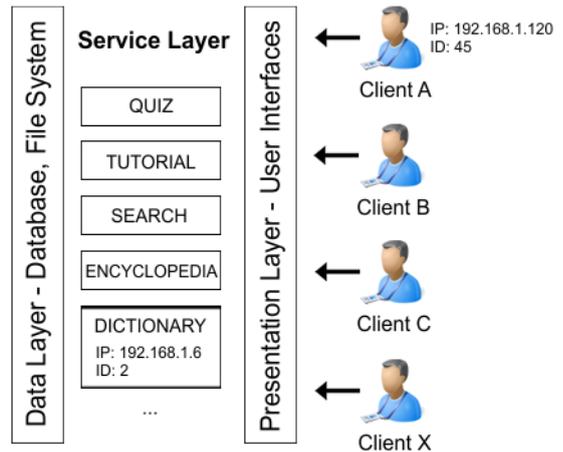

Figure 3.   E-Learning model

The "DICTIONARY" is a SOAP-based web service built using .NET language and has an IP equals to 192.168.1.6 and an ID equals to 2; while, "Client A" is a Java program that is going to request an operation from the "DICTIONARY" web service. Its IP is equal to 192.168.1.120 and its ID is equal to 45. The communication between "Client A" and the "DICTIONARY" web service is formatted using the ECL language and totally handled by the ECU of the ecosystem.

When tested, "Client A" invoked function "whatIs" over the "DICTIONARY" web service sending the parameter "word" with value "apple". The following is the request of "Client A" formatted in ECL.

```
<protocol>
    <sourceIP>192.168.1.120</sourceIP>
    <destinationIP>192.168.1.6</destinationIP>
    <sourceID>45</sourceID>
    <destinationID>2</destinationID>
    <functionInvoked>whatIs</functionInvoked>
    <functionParams>
        <name>word</name>
        <value>apple</value>
        <type>string</type>
    </functionParams>
    <stamp>12/4/2011 09:32:10PM</stamp>
    <version>1.0</version>
</protocol>
```

After the request has been received by the ECU, it is validated, parsed, and converted into a SOAP request and then forwarded to the SOAP adapter which, consecutively, forwards it to the "DICTIONARY" web service. Next is the request in SOAP protocol.

```
<?xml version="1.0"?>
<soap:Envelope
    xmlns:soap="http://www.w3.org/2001/12/soap-envelope"
    soap:encodingStyle="http://www.w3.org/2001/12/soap-encoding">
<soap:Body>
    <m:whatIs>
        <m:word>apple</m:word>
    </m:whatIs>
</soap:Body>
</soap:Envelope>
```

Once the "DICTIONARY" web service receives the SOAP request, it processes it and calls the internal function "whatIs" whose results are then formatted into a SOAP message and sent back to the SOAP adapter of the ECU. The returned SOAP response is shown below:

```
<?xml version="1.0"?>
<soap:Envelope
 xmlns:soap="http://www.w3.org/2001/12/soap-envelope"
 soap:encodingStyle="http://www.w3.org/2001/12/soap-
encoding">
<soap:Body>
 <m:WhatIsResponse>
  <m:meaning>fruit with red or yellow skin and sweet
taste</m:meaning>
 </m: WhatIsResponse >
</soap:Body>
</soap:Envelope>
```

Finally, the ECU converts the SOAP response into an ECL response and forwards it to "Client A". The ECL response is shown below:

```
<protocol>
    <sourceIP>192.168.1.6</sourceIP>
    <destinationIP>192.168.1.120</destinationIP>
    <sourceID>2</sourceID>
    <destinationID>45</destinationID>
    <returnValue>fruit with red or yellow skin and sweet
taste</functionReturnValue>
    <returnType>string</functionReturnType>
    <stamp>12/4/2011 09:32:13PM </stamp>
    <version>1.0</version>
</protocol>
```

## VIII. CONCLUSIONS & FUTURE WORK

This paper presented a language specification for exchanging structured data in digital ecosystems called ECL short for Ecosystem Communication Language. It relies on XML language to format its content mainly composed of metadata and payload. It additionally supports data encryption using the XML Encryption standard. The ECL is interpreted and processed by an internal unit called ECU short for Ecosystem Communication Unit. All together, they provide a transparent communication to all services connected to the ecosystem infrastructure by shielding the internal implementation and protocols of the existing heterogeneous distributed components. In addition, they deliver a standard data-path allowing a portable and interoperable interaction between the different entities of the ecosystem to send requests, and receive responses from each other, despite their incompatible architectures, platforms, and technologies.

As future work, the ECL is to support, in addition to encryption, digital signature for ensuring message's data integrity as well as its authenticity. This would allow the reliable exchange of ECL messages between connected parties regardless of the errors and noise present in the communication channels.

## IX. ACKNOWLEDGMENTS

This research was funded by the Lebanese Association for Computational Sciences (LACSC), Beirut, Lebanon under the "Digital Ecosystem Research Project – DERP2011".